\documentclass[aps,pre,twocolumn,groupedaddress,showpacs]{revtex4}
\usepackage{graphicx}
\usepackage{dcolumn}
\usepackage{bm}
\input epsf
\epsfclipon
\begin{document}
\title{Average path length in uncorrelated random networks with hidden variables}
\author{Agata Fronczak, Piotr Fronczak and Janusz A. Ho\l yst}
\affiliation{Faculty of Physics and Center of Excellence for
Complex Systems Research, Warsaw University of Technology,
Koszykowa 75, PL-00-662 Warsaw, Poland}
\date{\today}

\begin{abstract}
Analytic solution for the average path length in a large class of
uncorrelated random networks with hidden variables is found. We
apply the approach to classical random graphs of Erd\"{o}s and
R\'{e}nyi $(ER)$, evolving networks introduced by Barab\'{a}si and
Albert as well as random networks with asymptotic scale-free
connectivity distributions characterized by an arbitrary scaling
exponent $\alpha>2$. Our result for $2<\alpha<3$ shows that
structural properties of asymptotic scale-free networks including
numerous examples of real-world systems are even more intriguing
then ultra-small world behavior noticed in {\it pure} scale-free
structures and for large system sizes $N\rightarrow\infty$ there
is a saturation effect for the average path length.
\end{abstract}

\pacs{89.75.-k, 02.50.-r, 05.50.+q}

\maketitle

\par During the last few years random, evolving networks have become a
very popular research domain among physicists
\cite{0a,0b,BArev,2}. A lot of efforts were put into investigation
of such systems, in order to recognize their structure and to
analyze emerging complex properties. It was observed that despite
network diversity, most of real web-like systems share three
prominent structural features: small average path length ($APL$),
high clustering and {\it scale-free} (SF) degree distribution
\cite{0a,0b,BArev,2,14}. Several network topology generators have
been proposed to embody the fundamental characteristics
\cite{BA1,15,12,krapPRL2001,krapPRE2001,16,calPRL2002}.

\par To find out how the small-world property (i.e. small $APL$)
arises, the idea of shortcuts has been proposed by Watts and
Strogatz \cite{32}. To understand where the ubiquity of scale-free
distributions in real networks comes from, the concept of evolving
networks basing on preferential attachment has been introduced by
Barab\'{a}si and Albert \cite{BA1}. Recently Calderelli and
coworkers \cite{calPRL2002} have presented another mechanism that
accounts for origins of power-law connectivity distributions. It
is interesting that the mechanism is neither related to dynamical
properties nor to preferential attachment. Caldarelli et al. have
studied a simple static network model in which each vertex $i$ has
assigned a tag $h_{i}$ (fitness, hidden variable) randomly drawn
from a fixed probability distribution $\rho(h)$. In their fitness
model, edges are assigned to pairs of vertices with a given
connection probability $\widetilde{p}_{ij}$, depending on the
values of the tags $h_{i}$ and $h_{j}$ assigned at the edge end
points. Similar models have been also analyzed in several other
studies \cite{gohPRL2001,chuComb2002,sodPRE2002}.

\par A generalization of the above-mentioned network models has been
recently proposed by Bogu\~{n}\'{a} and Pastor-Satorras
\cite{bogPRE2003}. In the cited paper, the authors have argued
that such diverse networks like classical random graphs of
Erd\"{o}s and R\'{e}nyi $(ER)$, fitness model proposed by
Caldarelli et al. and even scale-free evolving networks introduced
by Barab\'{a}si and Albert $(BA)$ may be described by a common
formalism. Bogu\~{n}\'{a} and Pastor-Satorras have derived
analytical expressions for connectivity distributions $P(k)$ and
relations describing degree correlations in such networks as
functions of distributions of hidden variables $\rho(h)$ and the
probability of an edge establishment $\widetilde{p}_{ij}$. In this
paper we present an  analytical description of main topological
properties of the foregoing networks. We derive a general
theoretical formalism describing metric features (i.e. $APL$,
intervertex distance distribution) of random networks with hidden
variables, assuming that the connection probability scales as
$\widetilde{p}_{ij}\sim h_{i}h_{j}$ \cite{info1}. The last
assumption concerning the factorised form of $\widetilde{p}_{ij}$
translates into the absence of two-point correlations and applies
to a broad class of networks.

\par The issue of the small-world property is of great importance for
network studies.  The property directly affects such crucial
fields like information processing in different communication
systems (including the Internet) \cite{26,havPRL1,havPRL2,30},
disease or rumor transmission in social networks \cite{33,34,35}
as well as network designing and optimization
\cite{29,31,46,PRLoptimal}. Not long ago, there was a strong
belief that all the processes become more efficient when the mean
distance between network sites is smaller. Recently however, it
was shown that the small-world property may have an unfavorable
influence on such phenomena like synchronizability
\cite{PRLsynchro}.

\par Despite the universality and usefulness of the small-world
concept, except a few cases \cite{newPRL2000,szaPRE2002,
havPRLultra,dorNuc2003}, satisfactory calculations of the average
path length ($APL$) almost do not exists. Even in the case of {\it
aged} Erd\"{o}s - R\'{e}nyi graphs only a scaling relation (not an
exact formula) describing $APL$ is known $l_{ER}\sim\ln
N/\ln\langle k\rangle$ \cite{BArev}. In this paper we derive an
exact formula for the average distance $l_{ij}$ between any two
nodes $i$ and $j$ characterized by given values of hidden
variables $h_{i}$ and $h_{j}$. Averaging the quantity $l_{ij}$
over all pairs of vertices we obtain the average path length
characterizing the whole network. It is important to stress that
our formulas for $APL$ do not posses any free parameters,
therefore may be directly compared with results of computer
simulations. In this paper we have tested our analytic results
against numerical calculations performed for Erd\"{o}s - R\'{e}nyi
classical random graphs, $BA$ model and scale-free networks
$P(k)\sim k^{-\alpha}$ with arbitrary scaling exponent $\alpha$.
In all the cases we obtain a very good agreement between our
theoretical predictions and results of numerical investigation.

\par Let us start with the following lemma.
\newtheorem{tw}{Lemma}
\begin{tw}\label{tw1}
If $A_{1},A_{2},\dots,A_{n}$ are mutually independent events and
their probabilities fulfill relations $\forall_{i} P(A_{i})\leq
\varepsilon$ then
\begin{equation}
P(\bigcup_{i=1}^{n}A_{i})=1-\exp(-\sum_{i=1}^{n}P(A_{i}))-Q,
\end{equation}
where $0\leq Q<\sum_{j=0}^{n+1}
(n\varepsilon)^{j}/j!-(1+\varepsilon)^{n}$.
\end{tw}
{\bf Proof.} Using the method of inclusion and exclusion
\cite{feller} we get
\begin{eqnarray}\label{p1}
P(\bigcup_{i=1}^{n}A_{i})=\sum_{j=1}^{n}(-1)^{j+1}S(j),
\end{eqnarray}
with
\begin{eqnarray}\label{p2}
S(j)=\sum_{1 \leq i_{1} < i_{2} < \dots < i_{j} \leq n}^{n}
P(A_{i_{1}})P(A_{i_{2}})\dots P(A_{i_{j}}) \nonumber \\
=\frac{1}{j!}\left ( \sum_{i=1}^{n} P(A_{i}) \right )^{j}-Q_{j},
\end{eqnarray}
where $0\leq Q_{j}\leq \left( n^{j}/j! - \left( ^{n}_{j}
\right)\right)\varepsilon^{j}$. The term in bracket represents the total number
of redundant components occurring in the last line of (\ref{p2}). Neglecting
$Q_{j}$ it is easy to see that $(1-P(\cup A_{i}))$ corresponds to the first
$(n+1)$ terms in the MacLaurin expansion of $\exp(-\sum P(A_{i}))$. The effect
of higher-order terms in this expansion is smaller than
$R<(n\varepsilon)^{n+1}/(n+1)!$. It follows that the total error of (\ref{tw1})
may be estimated as $Q<\sum_{j=1}^{n}Q_{j}+R$. This completes the proof.

\par Let us notice that the terms $Q_{j}$ in (\ref{p2}) disappear when
one approximates multiple sums $\sum_{1 \leq i_{1} < i_{2} < \dots
< i_{j} \leq n}^{n}$ by corresponding multiple integrals. For
$\varepsilon = A/n\ll 1$ the error of the above assessment is less
then $A^{2}\exp(A)/n$ and may be dropped in the limit
$n\rightarrow \infty$.

\par At the moment we briefly repeat (after Ref. \cite{bogPRE2003}) the
main properties of random networks with hidden variables and
connection probability $\widetilde{p}_{ij}$ given by
\begin{equation}\label{pij1}
\widetilde{p}_{ij}=\frac{h_{i}h_{j}}{\beta},
\end{equation}
where $\beta$ is a certain constant. In the case of random
networks, where two-point correlations at the level of hidden
variables are absent we have
\begin{equation}\label{beta}
\beta=\langle h\rangle N,
\end{equation}
whereas in correlated BA networks the prefactor gains another
form. Bogu\~{n}\'{a} and Pastor-Satorras have shown that degree
distribution $P(k)$ in such uncorrelated networks is given by
\begin{equation}\label{Pk}
P(k)=\sum_{h}\frac{e^{-h}h^{k}}{k!}\rho(h),
\end{equation}
where $\rho(h)$ describes a distribution of hidden variables. The
above relation between both distributions $P(k)$ and $\rho(h)$
implies a relation between their moments
\begin{equation}\label{moments}
\langle h^{n}\rangle=\langle k(k-1)\dots(h-n+1)\rangle,
\end{equation}
and respectively
\begin{equation}\label{moments12}
\langle h\rangle=\langle k\rangle, \;\;\;\;\;\;\; \langle
h^{2}\rangle=\langle k(k-1)\rangle.
\end{equation}

\par With respect to our following calculations the relation (\ref{Pk})
requires a few comments. Firstly, let us note that for
$k\rightarrow \infty$ the Poisson-like propagator, that
accompanies the distribution $\rho(h)$ in the formula for $P(k)$,
is a sharply peaked function analogous to delta $\delta_{h,k}$.
For this reason, in the limit of large nodes degrees we obtain a
correspondence between the studied uncorrelated networks with
hidden variables and random graphs with a given degree sequence
(the so-called configuration model) \cite{newPRE2001}
\begin{equation}\label{Pkro}
P(k)\sim\rho(k).
\end{equation}

\par Another very important conclusion that comes from considerations
performed in Ref. \cite{bogPRE2003} and seems to affect our later
derivations is that we can not generate uncorrelated random
networks with power-law degree distribution $P(k)\sim k^{-\alpha}$
and the scaling exponent $2\leq\alpha<3$ by means of the
factorised probability (\ref{pij1}) (see also
\cite{masPRE2003,parkPRE2003}). The axiomatic definition of
probability requires $\widetilde{p}_{ij}\leq 1$, giving the
condition for the maximum value of the the hidden variable
$h_{max}\sim\sqrt{N}$. When we think about hidden variables as
about desired degrees (as sketched in the previous paragraph) the
condition for $k_{max}\simeq h_{max}$ is in contradiction to the
cut-off of the power-law degree distribution $k_{cut}\sim
N^{1/(\alpha-1)}$ \cite{info2} that allows for nodes with degrees
higher than $k_{max}$. For this reason, our formalism describing
metric properties of random uncorrelated networks should not work
well for SF networks with $2\leq\alpha<3$. In contrast to the
above discussion, we noticed that our analytical predictions are
consistent with numerical calculations performed for scale-free
networks with arbitrary scaling exponent $\alpha>2$. We suspect
that the unexpected conformity for networks with $2\leq\alpha<3$
may be related to the extreme small fraction of {\it bad pairs} of
nodes with large degrees that do not fulfill the condition
$\widetilde{p}_{ij}\leq 1$ (see {\it Appendix A}).

\par Now, we come back to the main subject of the paper, it means the
issue of the average path length in random networks. Let us
consider a walk of length $x$ crossing index-linked vertices
$\{i,v_{1},v_{2} \dots v_{(x-1)},j\}$. Because of the lack of
correlations the probability of such a walk is described by the
product $\widetilde{p}_{iv_{1}}\:\widetilde{p}_{v_{1}v_{2}}\:
\widetilde{p}_{v_{2}v_{3}}\dots \widetilde{p}_{v_{(x-1)}j}$, where
$\widetilde{p}_{ij}$  gives a connection probability between
vertices $i$ and $j$ (\ref{pij1}). At this stage it is important
to stress that the graph theory distinguishes {\it a walk} from
{\it a path} \cite{45}. A walk is just a sequence of vertices. The
only condition for such a sequence is that two successive nodes
must be the nearest neighbors. A walk is termed a path if all of
its vertices are distinct. In fact, we are interested in the
shortest paths. In order to do it, let us consider the situation
when there exists at least one walk of the length $x$ between the
vertices $i$ and $j$. If the walk(s) is(are) the shortest path(s)
$i$ and $j$ are exactly $x$-th neighbors otherwise they are closer
neighbors. In terms of statistical ensemble of random graphs
\cite{krzPRE2001} the probability $p_{ij}(x)$ of at least one walk
of the length $x$ between $i$ and $j$ expresses also the
probability that these nodes are neighbors of order not higher
than $x$. Thus, the probability that $i$ and $j$ are exactly
$x$-th neighbors is given by the difference
\begin{equation}\label{pijx*A}
p_{ij}^{*}(x)=p_{ij}(x)-p_{ij}(x-1).
\end{equation}

\par In order to write the formula for $p_{ij}(x)$ we take advantage of
the lemma (\ref{tw1})
\begin{eqnarray}\label{pijxA}
p_{ij}(x)=1-\exp[-\sum_{v_{1}=1}^{N}\dots
\sum_{v_{(x-1)}=1}^{N}\widetilde{p}_{iv_{1}}\dots
\widetilde{p}_{v_{(x-1)}j}],\;\;\;
\end{eqnarray}
where $N$ is the total number of vertices in a network. A sequence
of $(x+1)$ vertices $\{i,v_{1},v_{2}\dots ,v_{(x-1)},j\}$
beginning with $i$ and ending with $j$ corresponds to a single
event $A_{i}$ and the number of such events is given by
$n=N^{x-1}$. Putting (\ref{pij1}) into (\ref{pijxA}) and replacing
the sum over nodes indexes by the sum over the hidden variable
distribution $\rho(h)$ one gets
\begin{equation}\label{pijxB}
p_{ij}(x)=1-\exp\left[-\frac{h_{i}h_{j}}{\langle h^{2} \rangle
N}\left(\frac{\langle h^{2}\rangle N}{\beta}\right )^{x\;}\right].
\end{equation}
Due to (\ref{pijx*A}) the probability that both vertices are
exactly the $x$-th neighbors may be written as
\begin{equation}\label{pijx*B}
p_{ij}^{*}(x)=F(x-1)-F(x),
\end{equation}
where
\begin{equation}\label{Fx}
F(x)=\exp\left[-\frac{h_{i}h_{j}}{\langle h^{2} \rangle N
}\left(\frac{\langle h^{2}\rangle N}{\beta}\right )^{x\;}\right].
\end{equation}

\par The above calculations require a few comments. First of all, note
that the assumption underlying (\ref{tw1}) is the mutual
independence of all contributing events $A_{i}$. In fact, since
the same edge may participate in several $x-$walks there exist
correlations between these events. Nevertheless, it is easy to see
that the fraction of correlated walks is negligible for short
walks ($x\ll N$) that play the major role in random graphs showing
small-world behavior. It is also important to stress that our
formalism does not neglect loops.

\par Let us point out that having  relations (\ref{pijx*B}) and
(\ref{Fx}), describing the probability that the shortest distance
between two arbitrary nodes $i$ and $j$ equals $x$, one can find
almost all metric properties of studied networks \cite{afcond2}.
For example, averaging (\ref{pijx*B}) over all pairs of vertices
one obtains the intervertex distance distribution
$p(x)=\langle\langle p_{ij}^{*}(x)\rangle_{i}\rangle_{j}$. It is
also possible to calculate $z_{x}$ - the mean number of vertices a
certain distance $x$ away from a randomly chosen vertex $i$. The
quantity can be written as $z_{x}=\int p_{ij}^{*}(x)\rho(h_{j})N
dh_{j}$. Note that taking only the first two terms of power series
expansion of both exponential functions in (\ref{pijx*B}) and
making use of (\ref{pij1}) and (\ref{moments12}) one gets the
relationship $z_{x}=z_{1}(z_{2}/z_{1})^{x-1}=\langle k
\rangle(\langle k^{2}\rangle/\langle k\rangle-1)^{x-1}$ that was
obtained by Newman et al. \cite{newPRE2001} when assuming a
tree-like structure of random graphs with arbitrary degree
distribution.

\par Taking advantage of (\ref{pijx*B}) one can calculate the
expectation value for the average distance between $i$ and $j$
\begin{equation}\label{lijA}
l_{ij}(h_{i},h_{j})=\sum_{x=1}^{\infty}\:x\:p_{ij}^{*}(x)=\sum_{x=0}^{\infty}F(x).
\end{equation}
Notice that a walk may cross the same node several times thus the
largest possible walk length can be $x=\infty$. The Poisson
summation formula allows us to simplify the above sum (see {\it
Appendix B})
\begin{equation}\label{lij}
l_{ij}(h_{i},h_{j})=\frac{-\ln h_{i}h_{j}+\ln N+\ln\langle
h^{2}\rangle -\gamma}{\ln N+\ln\langle h^{2}\rangle-\ln\beta
}+\frac{1}{2},
\end{equation}
where $\gamma\simeq 0.5772$ is the Euler's constant. The average
intervertex distance for the whole network depends on a specified
distribution of hidden variables $\rho(h)$
\begin{equation}\label{lRG}
l=\frac{-2\langle\ln h\rangle+\ln N+\ln\langle h^{2}\rangle
-\gamma}{\ln N+\ln\langle h^{2}\rangle-\ln\beta }+\frac{1}{2}.
\end{equation}
We need to stress that both parameters $l_{ij}$ and $l$ diverge
when the argument of the logarithmic function in the denominator
of both expressions (\ref{lij}) and (\ref{lRG}) approaches one
i.e. $N \langle h^{2}\rangle/\beta=1$. Note, that substituting
(\ref{beta}) for $\beta$ in the last condition and then taking
advantage of (\ref{moments}) one recovers the well-known
estimation for percolation threshold $\langle k^{2}\rangle/\langle
k\rangle=2$ in undirected random networks with arbitrary degree
distribution \cite{havPRL1,molloy1,calPRL2000,golPRE2003} (see
{\it Appendix C}).

\par To test the formula (\ref{lRG}) we start with the well-known networks:
$ER$ classical random graphs, $BA$ model and scale-free networks.
The choice of these networks is not accidental. The models play an
important role in the network science. The $ER$ model was
historically the first one but it has been realized it is too
random to describe real networks. The most striking discrepancy
between $ER$ model and real networks appears when comparing degree
distributions. As mention at the beginning of the paper degree
distributions follow a power-law in most of real systems, whereas
classical random graphs exhibit Poisson degree distribution. It
was found that the most generic mechanism driving real networks
into scale-free structures is the linear preferential attachment.
The simplest model that incorporates the rule of preferential
attachment was introduced by Barab\'{a}si and Albert \cite{BA1}.
Other interesting mechanisms leading to scale-free networks were
proposed by Goh et al. \cite{gohPRL2001} and Caldarelli et al.
\cite{calPRL2002}. Goh and coworkers were the first who pointed
out that power-law connectivity distribution $P(k)$ may result
from the Zipf law applied to hidden variable distribution
$\rho(h)\sim h^{-\alpha}$. The concept of the Zipf law has been
next developed by Caldarelli et al. in their paper
\cite{calPRL2002}. In fact, the most important achievement of the
mention paper by Caldarelli et al. relates to a nontrivial
discovery that scale-free networks may be also obtained from
exponential distribution of fitnesses $\rho(h)\sim e^{-h}$. Since
however, the case of scale-free networks with exponentially
distributed fitnesses does not fulfill (\ref{pij1}), we do not
take it into account in this paper. In the present study, we
examine the case of scale-free networks with underlying scale-free
distributions of hidden variables.

\par Below we show that our formalism describing metric properties of
random networks may be successfully applied to all the above
listed network models.

\par
\begin{figure} \epsfxsize=6.4cm \epsfbox{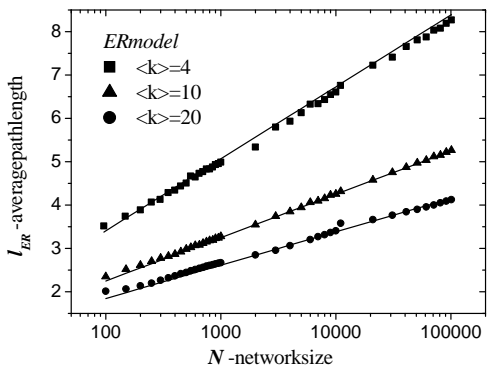}
\caption{The average path length $l_{ER}$ versus network size $N$
in $ER$ classical random graphs with $\langle k \rangle=4,10,20$.
The solid curves represent numerical prediction of
Eq.(\ref{lER}).} \label{figer}
\end{figure}

 {\it Classical $ER$ random graphs}. Note that the only way to
recover the Poisson degree distribution form the expression
(\ref{Pk}) is to assume
\begin{equation}\label{roER}
\rho_{ER}(h)=\delta_{\langle k \rangle,h}.
\end{equation}
Now, applying the distribution $\rho_{ER}(h)$ to (\ref{lRG}) we
get the formula for the average path length in classical random
graphs
\begin{equation}\label{lER}
l_{ER}=\frac{\ln N - \gamma}{\ln\langle k\rangle}+\frac{1}{2}.
\end{equation}
Until now only a rough estimation of the quantity has been known.
One has expected that the mean intervertex distance of the whole
ER graph scales with the number of nodes in the same way as the
network diameter. We remind that the diameter $d$ of a graph is
defined as the maximal shortest distance between any pair of
vertices and $d_{ER}=\ln N/\ln \langle k\rangle$ \cite{BArev}.
Fig.\ref{figer} shows the prediction of the equation (\ref{lER})
in comparison to the numerically calculated $APL$ in classical
random graphs.

\par
\begin{figure} \epsfxsize=6.6 cm \epsfbox{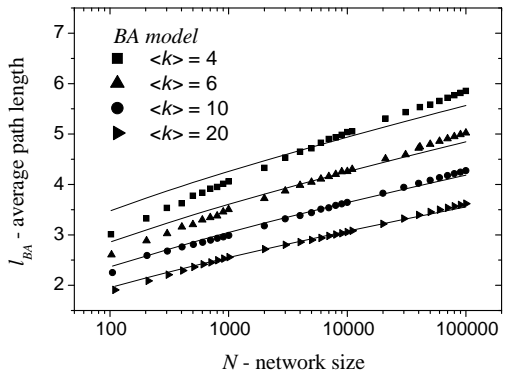}
\caption{Characteristic path length $l_{BA}$ versus network size
$N$ in $BA$ networks. Solid lines represent Eq.(\ref{lBA}).}
\label{figba}
\end{figure}

 {\it Scale-free $BA$ networks}. The basis of the $BA$ model is its
construction procedure \cite{BA1,BA2}. Two important ingredients
of the procedure are: the continuous network growth and the
preferential attachment. The network starts to grow from an
initial cluster of $m$ fully connected vertices. Each new node
that is added to the network creates $m$ links that connect it to
previously added nodes. The preferential attachment means that the
probability of a new link growing out of a vertex $i$ and ending
up in a vertex $j$ is given by
\begin{equation}\label{pijBA0}
\widetilde{p}_{ij}^{BA}=m\frac{k_{j}(t_{i})}{\sum_{l}k_{l}(t_{i})},
\end{equation}
where $k_{j}(t_{i})$ denotes the connectivity of a node $j$ at the
time $t_{i}$, when a new node $i$ is added to the network. Taking
into account the time evolution of nodes degree in $BA$ network
(i.e. putting $k_{j}(t_{i})=m\sqrt{t_{i}/t_{j}}$), the probability
of a link between $i$ and $j$ can be rewritten in the following
form
\begin{equation}\label{pijBA}
\widetilde{p}_{ij}^{BA}=\frac{m}{2}\frac{1}{\sqrt{t_{i}t_{j}}},
\end{equation}
that is equivalent to (\ref{pij1}) when assuming
$h_{i}=1/\sqrt{t_{i}}$, $h_{j}=1/\sqrt{t_{j}}$ and
$\beta_{BA}=2/m$. The distribution of hidden variables
$\rho_{BA}(h)$ in BA networks follows the relation
\begin{equation}
\rho_{BA}(h_{i})dh_{i}=\tilde{P}(t_i)dt_i,
\end{equation}
where $\tilde{P}(t_i)=1/N$ is the distribution of nodes attachment
times $t_i$ for a network of size $N$. After a simple algebra one
gets
\begin{equation}
\rho_{BA}(h)=\frac{2}{N}h^{-3},
\end{equation}
for $h=1/\sqrt{N},\dots,1$.
Now, it is simple to calculate the average distance (\ref{lij})
between any two nodes in BA networks
\begin{equation}\label{lijBA}
l_{ij}^{BA}(h_{i},h_{j})=\frac{-\ln(h_{i}h_{j})-\ln(m/2) -\gamma
}{\ln\ln N+\ln (m/2)}+\frac{3}{2}.
\end{equation}
Averaging (\ref{lijBA}) over all pairs of vertices one obtains
$APL$ characterizing the whole network
\begin{equation}\label{lBA}
l_{BA}=\frac{\ln N-\ln(m/2)-1-\gamma}{\ln\ln
N+\ln(m/2)}+\frac{3}{2}.
\end{equation}
Fig.\ref{figba} shows the $APL$ in $BA$ networks as a function of
the network size $N$ compared with the analytical formula
(\ref{lBA}). There is a visible discrepancy between the theory and
numerical results when $\langle k\rangle=2m=4$. The discrepancy
disappears when the network becomes denser i.e. when $\langle
k\rangle$ increases. The same effect will appear at
Fig.\ref{figAPLsf}, letting us deduce that for some reasons our
formalism works better for denser networks.

\par {\it Scale-free networks with arbitrary scaling exponent}. Let
us start with the well-known model of scale-free networks
introduced by Goh et al. ({\it Model $A$}) \cite{gohPRL2001} and
its certain modification proposed by Caldarelli et al. ({\it Model
$B$}) \cite{calPRL2002}. We show that both models $A$ and $B$
possess peculiar properties that make application of our
theoretical approach impossible. Next, we make use of a general
procedure described at the beginning of the paper to generate
uncorrelated networks with asymptotic power-law connectivity
distributions ({\it Model $C$}).

\par {\it Model $A$.} To construct the network one has to perform the
following steps: (i.) prepare a fixed number $N$ of vertices;
(ii.) assign fitness (hidden variable) $h_{i}=i^{-\tau}$, with
$0\leq\tau<1$, to every node $i=1,\dots,N$; (iii.) select two
vertices $i$ and $j$ with probabilities equal to normalized hidden
variables, $h_{i}/(\langle h\rangle N)$ and $h_{j}/ (\langle
h\rangle N)$, respectively, and add an edge between them unless
one already exists; (iv.) repeat previous steps until $mN$ edges
are made in the system. Goh and coworkers have showed that the
resulting network generated in accordance with the above procedure
exhibits asymptotic power-law degree distribution
\begin{equation}\label{pkGoh}
P(k)\sim k^{-\alpha},
\end{equation}
where
\begin{equation}\label{aGoh}
\alpha=1+1/\tau,
\end{equation}
that gives $2<\alpha<\infty$. Although in these networks
probability of a connection approximately factorizes (\ref{pij1})
\begin{equation}\label{pijGoh}
\widetilde{p}_{ij}^{A}=1-\left(1-\frac{h_{i}h{j}}{(\langle
h\rangle N)^{2}}\right)^{mN}\simeq \frac{h_{i}h_{j}}{\beta_{A}},
\end{equation}
where $\beta_{A}=\langle h\rangle^{2}N/m$, there is one important
feature  of the model. The non-analytic statement, included in the
step (iii.) of the construction procedure expressed as {\it add an
edge unless one already exists}, gives rise to uncontrolled
intervertex correlations both for large $m$ and small $\alpha<3$.

\par {\it Model $B$.} Caldarelli and coworkers have modified the
original model introduced by Goh et al. by assigning to nodes
random fitnesses $h_{i}$ taken from a given distribution
$\rho(h)\sim h^{-\alpha}$, instead of deterministic values
$h_{i}=i^{-\tau}$. They also assumed a modified edge establishment
process: for every pair of vertices $i$ and $j$ a link was drawn
with probability (\ref{pij1}), where $\beta_{B}=(h_{max})^{2}$.
Although the foregoing value of $\beta_{B}$ assures us of
$\widetilde{p}_{ij}^{B}<1$, it is strongly overestimated and makes
resulting networks very sparse with a large content of isolated
nodes \cite{info3}.

\begin{figure} \epsfxsize=6.8 cm \epsfbox{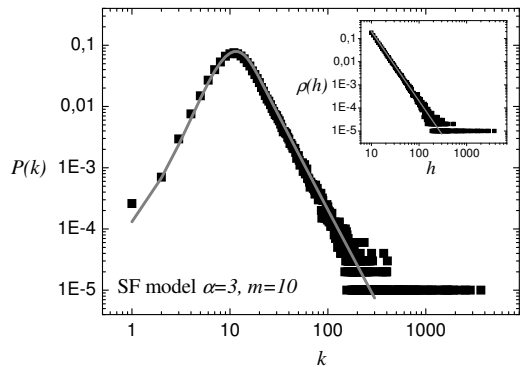}
\caption{{\it Model C.}  Degree distribution $P(k)$ (the main
layer) for random network with underlying hidden variable
distribution given by a power-law (the inset). Scatter data
represent results of numerical calculations, whereas solid curves
express formulas (\ref{Pk}) and (\ref{roSF}), respectively for
$P(k)$ and $\rho(h)$.} \label{figPkEF}
\end{figure}

\par {\it Model $C$.} In order to avoid features incorporated in
both models $A$ and $B$, we have generated networks possessing
asymptotic scale-free behavior for $k\gg 1$ coming out of
power-law distributions of hidden variables
\begin{equation}\label{roSF}
\rho(h)=\frac{(\alpha-1)m^{(\alpha-1)}}{h^{\alpha}},
\end{equation}
for $h=m,\dots,h_{max}$, where $h_{max}\simeq mN^{1/(\alpha-1)}$
(see \cite{info2}) and connection probability given by
(\ref{pij1}) and (\ref{beta}). A typical behavior of connectivity
distribution $P(k)$ for networks generated in accordance with this
procedure is presented at Fig.\ref{figPkEF}. Note that for $k>m$
the connectivity distribution is well described by the power law
$P(k)\sim k^{-\alpha}$ (\ref{Pkro}).

\begin{figure} \epsfxsize=6.8cm \epsfbox{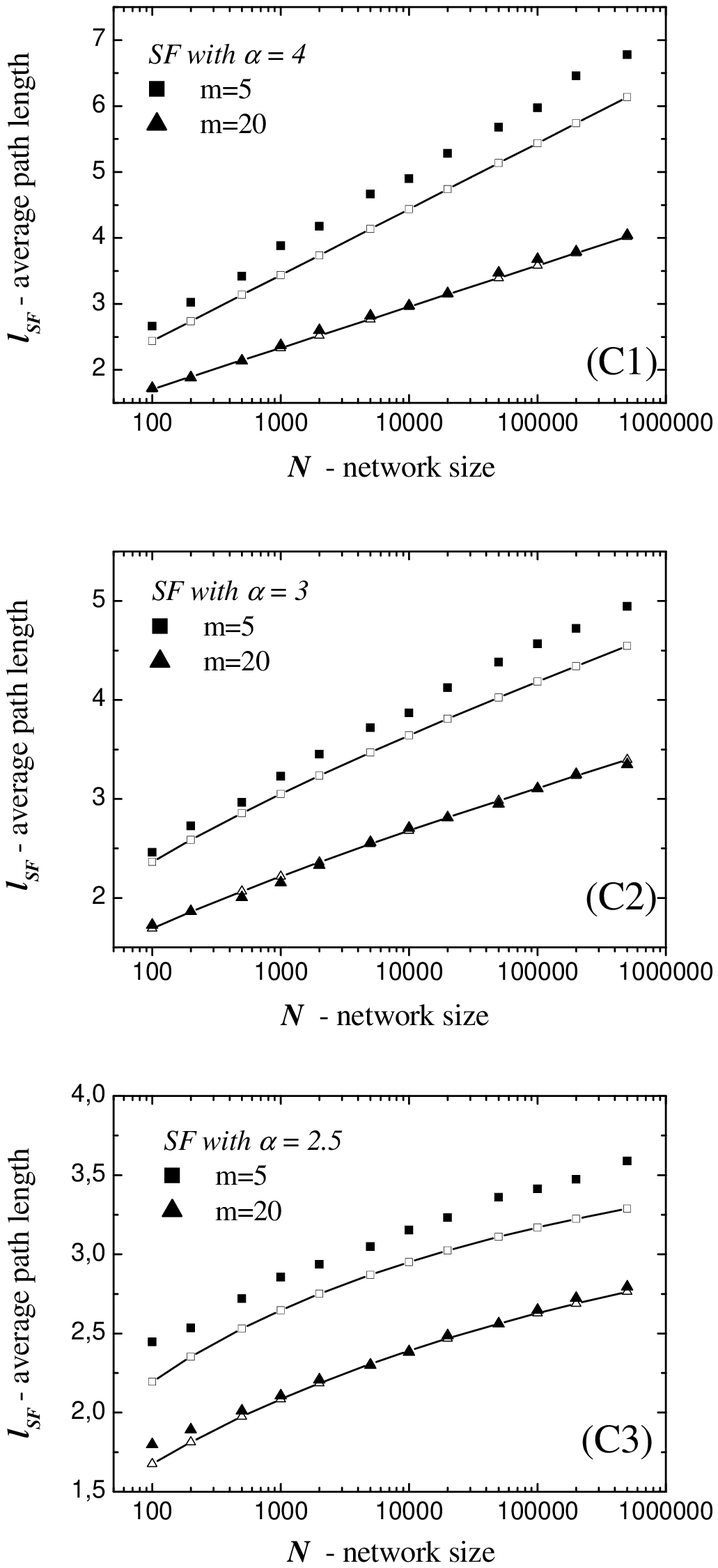}
\caption{{\it Model C.} The average path length versus network
size $N$ for $\alpha=4$ (C1), $\alpha=3$ (C2) and $\alpha=2.5$
(C3). The scatter data represent numerical calculations. Solid
curves with open squares in the case of $m=5$ (open circles in the
case of $m=20$) express analytical predictions of Eqs.
(\ref{APLa4}), (\ref{APLa3}) and (\ref{APLa25}) respectively for
(C1),(C2) and (C3).} \label{figAPLsf}
\end{figure}

\par Applying the distribution (\ref{roSF}) to the formula (\ref{lRG})
one obtains
\begin{itemize}
\item for $\alpha>3$
\begin{equation}\label{APLa4}
l^{\alpha>3}=\frac{\ln N+\ln(\frac{\alpha-1}{\alpha-3})-
\frac{2}{\alpha-1}-\gamma}{\ln(\frac{\alpha-2}{\alpha-3})+\ln m
}+\frac{1}{2},
\end{equation}
\item for $\alpha=3$
\begin{equation}\label{APLa3}
l^{\alpha=3}=\frac{\ln N-\ln(\frac{m}{2})-1-\gamma}{\ln\ln
N+\ln(\frac{m}{2})}+\frac{3}{2},
\end{equation}
\item for $2<\alpha<3$
\begin{equation}\label{APLa25}
l^{\alpha<3}=\frac{(\frac{2}{\alpha-1})\ln N
+\ln(\frac{\alpha-1}{3-\alpha})-(\frac{2}{\alpha-1})-\gamma}
{(\frac{3-\alpha}{\alpha-1})\ln N+\ln
(\frac{\alpha-2}{3-\alpha})+\ln m}+\frac{1}{2}.
\end{equation}
\end{itemize}
Fig.\ref{figAPLsf} shows predictions of the above equations in
comparison with numerically calculated shortest paths. We would
like to stress that regardless of the value of $\alpha$, for
denser networks (with higher values of parameter $m$), one can
observe an excellent agreement between our theory and numerical
results.

\par Summarizing, depending on the value of scaling exponent
$\alpha$ one can distinguish three scaling regions for the average
path length in scale-free networks. In the limit of large systems
$N\to\infty$, $APL$ scales with network size according to
relations
\begin{itemize}
\item for $\alpha>3$
\begin{equation}\label{APLa4N}
l^{\alpha>3}\sim \ln N,
\end{equation}
\item for $\alpha=3$
\begin{equation}\label{APLa3N}
l^{\alpha=3}\sim \frac{\ln N}{\ln\ln N},
\end{equation}
\item for $2<\alpha<3$
\begin{equation}\label{APLa25N}
l^{\alpha<3} = \frac{2}{3-\alpha}+\frac{1}{2}.
\end{equation}
\end{itemize}
Note that although the results for $\alpha\geq 3$ are consistent
with estimations obtained by other authors
\cite{havPRLultra,dorNuc2003}, the case of $2<\alpha<3$ is
different. In opposite to previous estimations signaling the
double logarithmic dependence $l^{\alpha<3}\sim\ln\ln N$, our
calculations for the same range of $\alpha$ predict that there is
a saturation effect for the mean path length in large networks.
Since the assumption underlying estimations leading to double
logarithmic dependence in $APL$ was a {\it pure} scale-free
behavior of degree distribution, we suspect that this discrepancy
may result from {\it ambiguous} behavior of $P(k)$ in our model.
Let us note that in our  model $C$ there is a relatively small
number of nodes with small degrees $k$ (see Fig.\ref{figPkEF}).
Since distances between such nodes are usually very large in
comparison to distanced between nodes with higher degrees, thus
their absence may lead to the domination of the $APL$ parameter by
distances between the population of nodes characterized by medium
degrees. Our result shows that for $2<\alpha<3$ structural
properties of asymptotic scale-free networks including numerous
examples of real-world networks may be even more intriguing then
ultra-small world behavior reported for pure scale-free systems.

\par To conclude, in this paper we have presented theoretical approach
for metric properties of uncorrelated random networks with hidden
variables. We have derived a formula for probability
$p^{*}_{ij}(x)$ (\ref{pijx*B}) that the shortest distance between
two arbitrary nodes $i$ and $j$ equals $x$. We have shown that
given $p^{*}_{ij}(x)$ one can find every structural characteristic
of the studied networks. In particular, we have applied our
approach to calculate exact expression for the average path length
(\ref{lRG}) in such networks. We have shown that our formalism may
be successfully applied to such diverse networks like classical
random graphs of Erd\"{o}s and R\'{e}nyi, evolving networks
introduced by Barab\'{a}si and Albert as well as random networks
with asymptotic scale-free connectivity distributions
characterized by arbitrary scaling exponent $\alpha>2$. In all the
studied cases we noticed a very good agreement between our
theoretical predictions and results of numerical investigation.

\par {\it Acknowledgments}. First of all, we wish to thank to anonymous
Referee for critical comments which helped us to refine our paper.
In the preliminary version of the paper \cite{afcond1} we have
used a {\it node degree} notation, that was not accurate enough
with reference to the analyzed problem. Following the comments
given by the anonymous Referee we have reformulated our approach
in a language of {\it hidden variables}. We are also thankful to
Sergei Dorogovtsev  for his critical comments to an earlier draft
of this paper. One of us (AF) acknowledge The State Committee for
Scientific Research in Poland for support under grant No. $2 P03B
013 23$.

\par {\it Appendix A}. The condition $\widetilde{p}_{ij}\leq 1$
(\ref{pij1}) is not fulfilled for pairs of vertices $i$ and $j$
possessing large hidden variables (or desired degrees) $h_{i}$ and
$h_{j}$. To justify our calculations, we have to assure ourselves
that the fraction of such pairs is very small
\begin{eqnarray}\label{rgcond}
\int_{h_{min}}^{h_{max}}\rho(h_{j})\int_{\widetilde{p}_{ij}\beta
/h_{j}}^{h_{max}} \rho(h_{i})dh_{i}dh_{j}\ll 1.
\end{eqnarray}
Using the Chebyshev's inequality \cite{feller} and solving
(\ref{rgcond}) with respect to $\widetilde{p}_{ij}\leq 1$ one gets
\begin{eqnarray}
\frac{\langle h^{2}\rangle}{\langle h\rangle^{2}}(\langle h^{2}
\rangle-\langle h\rangle^{2})\ll N^{2},
\end{eqnarray}
where we assumed $\beta=\langle h\rangle N$. It can be shown that
every network that is considered in this paper fulfill the
condition.

\par {\it Appendix B}. The Poisson summation formula states
\begin{eqnarray}\label{poissonform}
\sum_{x=0}^{\infty}F(x)=\frac{1}{2}F(0)+\;\;\;\;\;\;\;\;\;\;\;\;\;\;
\;\;\;\;\;\;\;\;\;\;\;\;\;\;
\;\;\;\;\;\;\;\;\;\;\;\;\;\;\;\;\;\;\;\;\;\;\;\;\;\;
\\\;\;\;\;\;\;\;\;\;\;\;\;\;\;\;
\int_{0}^{\infty}F(x)dx
+2\sum_{n=1}^{\infty}\left(\int_{0}^{\infty}F(x)\cos(2n\pi
x)dx\right)\nonumber.
\end{eqnarray}
Applying the formula to (\ref{lijA})
\begin{eqnarray}
l_{ij}(h_{i}h_{j})=\sum_{x=0}^{\infty}\exp\left[-\frac{h_{i}h_{j}}{\langle
h^{2} \rangle N}\left(\frac{\langle h^{2}\rangle N}{\beta}\right
)^{x\;}\right]
\end{eqnarray}
one realizes that in most of cases
\begin{eqnarray}
\frac{h_{i}h_{j}}{\langle h^{2}\rangle N}\simeq 0
\end{eqnarray}
that gives $F(0)=1$. One can also find that
\begin{eqnarray}
\int_{0}^{\infty} F(x)dx = -Ei\left(-\frac{h_{i}h_{j}}{\langle
h^{2}\rangle N}\right)/\ln\left(\frac{\langle h^{2}\rangle N
}{\beta}\right),
\end{eqnarray}
where $Ei(y)$ is the exponential integral function that for
negative arguments is given by $Ei(-y)=\gamma+\ln y$ \cite{ryzyk},
where $\gamma\simeq 0.5772$ is the Euler's constant. Finally, it
is easy to see that owing to the generalized mean value theorem
every integral in the last term of the summation formula
(\ref{poissonform}) is equal to zero. It follows that the equation
for the APL between $i$ and $j$ is given by (\ref{lij}).

\par {\it Appendix C}. Note that, using additional
assumptions one can simply reformulate both formulas (\ref{lij})
and (\ref{lRG}) as well as the whole formalism in terms of node's
degrees instead of hidden variables. For more details see
\cite{afcond1}.




\begin{thebibliography}{7}

\bibitem{0a} S.Bornholdt and H.G.Schuster, {\it Handbook of Graphs and
networks}, Wiley-Vch (2002).
\bibitem{0b} S.N. Dorogovtsev and J.F.F.Mendes, {\it Evolution of
Networks}, Oxford Univ.Press (2003).
\bibitem{BArev} R.Albert and A.L.Barab\'asi, Rev. Mod. Phys. {\bf 74} 47 (2002).
\bibitem{2} S.N.Dorogovtshev and J.F.F.Mendes, Adv.Phys. {\bf 51} 1079 (2002).
\bibitem{14} S.H.Strogatz, Nature {\bf 410} 268 (2001).
\bibitem{BA1} A.L.Barab\'asi and R.Albert, Science {\bf 286}, 509 (1999).
\bibitem{15} R.Albert and A.L.Barab\'asi, Phys. Rev. Lett. {\bf 85} 5234 (2000).
\bibitem{12} S.N.Dorogovtsev et al., Phys. Rev. Lett. {\bf 85} 4633 (2000).
\bibitem{krapPRL2001} P.L.Krapivsky et al., Phys. Rev. Lett. {\bf 86} 5401 (2001).
\bibitem{krapPRE2001} P.L. Krapivsky and S. Redner, Phys. Rev. E {\bf 63} 066123 (2001).
\bibitem{16} K.Klemm and V.M.Egu\'iluz, Phys. Rev. E {\bf 65} 036123 (2002).
\bibitem{calPRL2002} G. Caldarelli et al., Phys. Rev. Lett. {\bf 89}, 258702 (2002).
\bibitem{32} D.J.Watts and S.H.Strogatz, Nature {\bf 393} 440 (1998).
\bibitem{gohPRL2001} K.-I. Goh et al., Phys. Rev. Lett. {\bf 87}, 278701 (2001).
\bibitem{chuComb2002} F. Chung and L. Lu, Annals of Combinatorics {\bf 6}, 125 (2002).
\bibitem{sodPRE2002} B. S\"{o}derberg, Phys. Rev. E {\bf 66}, 066121 (2002).
\bibitem{bogPRE2003} M. Bogu\~{n}\'{a} and R. Pastor-Satorras, Phys. Rev. E {\bf 68}, 036112 (2003).
\bibitem{info1} In fact, all the derivations presented in this
paper may be simply reformulated when assume a general factorised
form of $p_{ij}=f(h_{i})g(h_{j})$, where $g$ and $f$ denote
arbitrary functions.
\bibitem{26} R.Albert et al., Nature {\bf 406}, 378 (2000).
\bibitem{havPRL1} R. Cohen et al., Phys. Rev. Lett. {\bf 85}, 4626 (2000).
\bibitem{havPRL2} R. Cohen et al., Phys. Rev. Lett. {\bf 86}, 3682 (2001).
\bibitem{30} R.Pastor-Satorras et al., Phys. Rev. Lett {\bf 87} 258701 (2001).
\bibitem{33} V.M.Egu\'{i}luz and K.Klemm, Phys. Rev. Lett. {\bf 89} 108701 (2002).
\bibitem{34} R.Pastor-Satorras and A.Vespignani, Phys. Rev. Lett. {\bf 86} 3200 (2001).
\bibitem{35} Z.Dezs\H{o} and A.L.Barab\'{a}si, Phys. Rev. E {\bf 65} 055103 (2002).
\bibitem{29} L.A.Adamic et al., Phys. Rev. E {\bf 64} 046135 (2001).
\bibitem{31} B.J.Kim et al., Phys. Rev. E {\bf 65} 027103 (2002).
\bibitem{46} S.Valverde et al., cond-mat/0204344 (2002).
\bibitem{PRLoptimal} L.A. Braunstein et al., Phys. Rev. Lett. {\bf 91} 168701 (2003).
\bibitem{PRLsynchro} T. Nishikawa et al., Phys. Rev. Lett. {\bf 91} 014101 (2003).
\bibitem{newPRL2000} M.E.J. Newman et al., Phys. Rev. Lett. {\bf 84}, 3201 (2000).
\bibitem{szaPRE2002} G. Szab\'{o} et al., Phys. Rev. E {\bf 66}, 026101 (2002).
\bibitem{havPRLultra} R. Cohen and S. Havlin, Phys. Rev. Lett. {\bf 90} 058701 (2003).
\bibitem{dorNuc2003} S.N. Dorogovtsev et al., Nucl. Phys. B {\bf 653}, 307 (2003).
\bibitem{feller} W.Feller, {\it An Introduction to Probability Theory and its Applications}, John Wiley and Sons (1968).
\bibitem{newPRE2001} M.E.J. Newman et al., Phys. Rev. E {\bf 64}, 026118 (2001).
\bibitem{masPRE2003} S. Maslow et al., cond-mat/0205379.
\bibitem{parkPRE2003} J. Park and M.E.J. Newman, Phys. Rev. E {\bf 68}, 026112 (2003).
\bibitem{info2} In a finite network, the cut-off of degree distribution
$P(k)\sim k^{-\alpha}$ may be estimated from
$\int_{k_{cut}}^{\infty}P(k)=1/N$ yielding $k_{cut}\sim
N^{1/(\alpha-1)}$.
\bibitem{45} R.J.Wilson, {\it Intr. to Graph Theory}, Longman (1985).
\bibitem{krzPRE2001} Z. Burda et al., Phys. Rev. E {\bf 64}, 046118 (2001).
\bibitem{afcond2} A. Fronczak et al., {\it How to calculate the main characteristics of random graphs - a new approach}, cond-mat/0308629.
\bibitem{molloy1} M. Molloy and B. Reed, Ran. Struct. and Algor. {\bf 6}, 161 (1995).
\bibitem{calPRL2000} D.S. Callaway et al., Phys. Rev. Lett. {\bf 85} 5468 (2002).
\bibitem{golPRE2003} A.V. Goltsev et al., Phys. Rev. E {\bf 67} 026123 (2003).
\bibitem{BA2} A.L.Barab\'asi et al., Physica A {\bf 272} 173 (1999).
\bibitem{info3} We have numerically checked that in the case of scaling exponents $\alpha=2.5$
and network size $N=10^{4}$ (see Fig.2 in \cite{calPRL2002}) the
amount of isolated vertices approaches $90\%$.
\bibitem{afcond1} A. Fronczak et al., {\it Average path lenght in random networks}, cond-mat/0212230.
\bibitem{ryzyk} I.S. Gradshteyn et al., {\it Table of Integrals, Series, and Products}, Academic Press (2000).

\end{thebibliography}
\end{document}